\documentclass[pra,twocolumn,showpacs,preprintnumbers,amsmath,amssymb,%
nofootinbib,floatfix]{revtex4}
\usepackage{graphicx}
\usepackage{dcolumn}
\usepackage{bm}
\usepackage{hyperref}

\newcommand{\Eq}[1]{Eq.\@ (\ref{#1})}
\newcommand{\Eqs}[1]{Eqs.\@ (\ref{#1})}
\newcommand{\Fig}[1]{Fig.\@ \ref{#1}}
\newcommand{\Figs}[1]{Figs.\@ \ref{#1}}
\newcommand{\Ref}[1]{Ref.\@ \cite{#1}}
\newcommand{\Refs}[1]{Refs.\@ \cite{#1}}
\newcommand{\Sec}[1]{Sec.\@ \ref{#1}}

\newcommand{\ave}[1]{\langle #1\rangle}

\newcommand{\vek}[1]{\bm{\mathrm{#1}}}
\newcommand{\nablav}{\bm{\nabla}}
\newcommand{\pv}{\vek{p}}
\newcommand{\rv}{\vek{r}}
\newcommand{\vv}{\vek{v}}

\newcommand{\dip}{\mathit{dip}}
\newcommand{\oct}{\mathit{oct}}
\newcommand{\bend}{\mathit{bend}}
\newcommand{\slosh}{\mathit{slosh}}
\newcommand{\eq}{\mathit{eq}}
\newcommand{\feq}{f_\eq}
\newcommand{\fbareq}{\bar{f}_\eq}
\newcommand{\ho}{\mathit{ho}}
\newcommand{\rad}{\perp}

\newcommand{\cpr}{C}
\newcommand{\cse}{C^{\prime}}
\newcommand{\Vexp}{V_{\mathit{exp}}}
\newcommand{\Vsix}{V_{\mathit{6th}}}
\newcommand{\Vtrap}{V_T}
\newcommand{\Vhat}{\hat{V}}
\newcommand{\Vhatslo}{\Vhat_\slosh}

\newcommand{\wrad}{\omega_\rad}
\newcommand{\wslo}{\omega_\slosh}
\newcommand{\wx}{\omega_x}
\newcommand{\wy}{\omega_y}
\newcommand{\wz}{\omega_z}
\newcommand{\wzmag}{\omega_{z,\mathit{mag}}}
\newcommand{\wzopt}{\omega_{z,\mathit{opt}}}

\newcommand{\coll}{\mathit{coll}}
\newcommand{\deltaU}{{\delta U}}
\newcommand{\trans}{\mathit{trans}}

\newcommand{\fst}{\mathit{1st}}
\newcommand{\trd}{\mathit{3rd}}

\newcommand{\dgamma}{\frac{d^3rd^3p}{(2\pi)^3}}
\newcommand{\dgammatext}{d^3rd^3p/(2\pi)^3}
\newcommand{\rrad}{r_\rad}

\DeclareMathOperator{\Imag}{Im}

\renewcommand{\Im}{\Imag}

\begin{document}

\title{Trap anharmonicity and sloshing mode of a Fermi gas}

\author{Pierre-Alexandre Pantel}
\affiliation{Universit{\'e} de Lyon, F-69622 Lyon, France;
  Univ. Lyon 1, Villeurbanne;
  CNRS/IN2P3, UMR5822, IPNL}
\author{Dany Davesne}
\affiliation{Universit{\'e} de Lyon, F-69622 Lyon, France;
  Univ. Lyon 1, Villeurbanne;
  CNRS/IN2P3, UMR5822, IPNL}
\author{Silvia Chiacchiera}
\affiliation{Centro de F{\'i}sica Computacional, Department of Physics,
University of Coimbra, P-3004-516 Coimbra, Portugal}
\author{Michael Urban}
\affiliation{Institut de Physique Nucl{\'e}aire, CNRS/IN2P3 and
  Universit\'e Paris-Sud 11, 91406 Orsay Cedex, France}
\begin{abstract}
For a gas trapped in a harmonic potential, the sloshing (or Kohn) mode
is undamped and its frequency coincides with the trap frequency,
independently of the statistics, interaction and temperature of the
gas. However, experimental trap potentials have usually Gaussian shape
and anharmonicity effects appear as the temperature and, in the case
of Fermions, the filling of the trap are increased. We study the
sloshing mode of a degenerate Fermi gas in an anharmonic trap within
the Boltzmann equation, including in-medium effects in both the
transport and collision terms. The calculated frequency shifts and
damping rates of the sloshing mode due to the trap anharmonicity are
in satisfactory agreement with the available experimental data. We
also discuss higher-order dipole, octupole, and bending modes and show
that the damping of the sloshing mode is caused by its coupling to
these modes.
\end{abstract}
\pacs{67.85.Lm}
\maketitle

\section{\label{sec:intro}Introduction}
In a couple of experiments, the measurement of the frequencies of
collective modes in trapped Fermi gases revealed a lot of interesting
information on the equation of state, the validity of superfluid
hydrodynamics, and the superfluid-normal phase transition
\cite{KinastHemmer2004, BartensteinAltmeyer2004, KinastTurlapov2004,
  AltmeyerRiedl2007a, AltmeyerRiedl2007b}. More recently, the
transition from the hydrodynamic to the collisionless regime in the
normal phase was also studied \cite{Wright2007,Riedl2008}. However,
since the frequencies of the collective modes depend on the trap
frequencies, a precise knowledge of the latter is required for a
meaningful interpretation of the collective-mode data.

A possibility to determine the trap frequency with high precision is
the measurement of the frequency of the sloshing mode
\cite{AltmeyerRiedl2007a}, which is an oscillation of the center of
mass of the trapped atom cloud. In a harmonic trap, this oscillation
(also called Kohn mode) is undamped and its frequency coincides
exactly with the trap frequency in the corresponding direction,
independently of the number of atoms $N$, or their interaction, their
temperature $T$ etc. This is a consequence of the Kohn theorem
\cite{Kohn,Brey} and follows from the fact that the center-of-mass
oscillation decouples completely from the internal dynamics of the gas
if the interaction is translationally invariant and the external
potential is harmonic \cite{Dobson}.

However, in practice the trap potential is never exactly harmonic. In
optical dipole traps \cite{GrimmWeidemueller00}, the potential is
typically Gaussian (corresponding to the intensity profile of the
laser beam). Because of this anharmonicity, the frequency of the
sloshing mode is shifted, and the shift depends on the system
parameters such as $N$, $T$, the scattering length $a$ characterizing
the interaction strength, etc. Furthermore, the sloshing mode is no
longer undamped.

The anharmonicity of the trap complicates considerably the analysis of
collective-mode experiments. In \Ref{Riedl2008} the measured
frequencies were corrected for the anharmonicity effects by giving
them in units of the measured frequency of the sloshing mode. However,
it is clear that the damping rates of the modes cannot be corrected in
this way. In \Ref{WuZhang2012_2d}, the damping rate of the sloshing mode
was used to estimate the increase of the damping rate of other modes
due to the anharmonicity. Both corrections are ad-hoc prescriptions
without rigorous justification. It is therefore strongly desirable to
get a better understanding of the anharmonicity effects on the
sloshing mode.

The aim of the present paper is to describe theoretically the
frequency shift and the damping rate of the sloshing mode in an
anharmonic trap. We will compare our results with the experimental
data available from the Innsbruck group
\cite{Riedl2008,Sidorenkov2012} and with the numerical results by Wu
and Zhang \cite{WuZhang2012_3d}. We will also discuss in detail the
damping mechanism of the sloshing mode. In an anharmonic potential,
the center-of-mass motion is no longer decoupled from the internal
degrees of freedom of the cloud. We will see that the damping of the
transverse sloshing mode is a consequence of its coupling to other
damped collective modes, in particular to the radial dipole mode and
the bending mode.

The framework of our study is the Boltzmann equation, including mean field
\cite{Chiacchiera2009} and in-medium cross-section
\cite{BruunSmith2007,Riedl2008,Chiacchiera2009}. Especially the mean field
is expected to be important in the present context, because it
can have a sizable effect on the density profile, i.e., on how far the
cloud extends into the anharmonic region of the trap potential. The
Boltzmann equation is solved approximately with the help of the
phase-space moments method. This method, when extended beyond the
lowest order, has proven to be in very good quantitative agreement
with the results of a full numerical simulation \cite{Lepers2010}. It
has also been quite successful for the description of the experimental
results for the frequency and damping rate of the radial quadrupole
mode \cite{Chiacchiera2011}. In the present case of the sloshing mode,
we include phase-space moments of first and third order.

The paper is organized as follows. The general formalism is briefly
presented in \Sec{sec:formalism}. Then we specify our model for the
experimental trap potential in \Sec{sec:potential}. In
\Sec{sec:firstorder}, we give a formula for the frequency shift of the
sloshing mode within the first-order moments method. Then we extend
the ansatz to third order in \Sec{sec:thirdorder}. The physical
contents of the extended ansatz and the numerical results are
discussed. In \Sec{sec:cubic} the third-order ansatz is used to
describe also the radial dipole, radial octupole, and bending
modes. Finally, in \Sec{sec:cl}, we will conclude.

Throughout the paper, we use units with $\hbar=k_B=1$.

\section{\label{sec:formalism}Summary of the formalism}
In this section, we give a short summary of the formalisms of
\Refs{Chiacchiera2009,Chiacchiera2011}. More details can be found
there.
\subsection{Linearized Boltzmann equation with in-medium effects}
We consider a balanced two-component ($N_\uparrow=N_\downarrow=N/2$)
Fermi gas of atoms with mass $m$ and interspecies attractive
interaction (scattering length $a<0$), trapped in a potential
$\Vtrap(\rv)$. The framework we use to describe the collective
dynamics of the system in the normal-fluid phase is the Boltzmann
equation. We assume that the two components move in phase, so that
only one distribution function $f = f_\uparrow=f_\downarrow$ is
needed. It is normalized to $\int \dgammatext f = N/2$, and
expectation values of one-body operators are given by
\begin{equation}
\ave{q}(t) = \frac{2}{N} \int \dgamma f(\rv,\pv,t) q(\rv,\pv)\,.
\end{equation}
As in \Ref{Chiacchiera2009}, we include in-medium effects in both the
transport and the collision parts of the Boltzmann equation: a
mean field like potential $U$ and the in-medium modified cross-section
$d\sigma/d\Omega$, respectively.

Within the Thomas-Fermi or local-density approximation (LDA), the
equilibrium distribution function reads
\begin{equation}
  \feq(\rv,\pv) =
\frac{1}{e^{\beta[\frac{\pv^2}{2 m}+\Vtrap(\rv)+U_\eq(\rv)-\mu_0]} + 1},
\end{equation}
where $U_\eq$ is the mean field in equilibrium, $\mu_0$ is the
chemical potential, and $\beta=1/T$ is the inverse temperature. As in
\Ref{Chiacchiera2009}, we obtain $U_\eq$ from the single-particle
self-energy in ladder approximation, evaluated at the Fermi level. In
the weak coupling limit, this reduces to the Hartree term
$U_\mathrm{Hartree} = 4\pi a\rho/m$, where $\rho = \int d^3p/(2\pi)^3
f$ is the density per spin state, and it remains finite for all
interaction strengths up to the unitary limit, $a \to -\infty$. In
equilibrium, the main effect of the mean field is to enhance the
density in the center of the trap, as shown in Fig.\@ 3 of
\Ref{Chiacchiera2009}. 

For the study of collective oscillations, it is sufficient to consider
small deviations from equilibrium and to linearize the Boltzmann
equation with respect to $\delta f = f-\feq$. If we write the
variation of the distribution function in the form \cite{Landau10}
\begin{equation}
\delta f(\rv,\pv,t) = \feq \fbareq\Phi(\rv,\pv,t)\,,
\end{equation}
with $\fbareq = 1-\feq$, the linearized Boltzmann equation can be written
as
\begin{multline}
  \feq\fbareq \Big(\dot{\Phi}
  +\Big\{\Phi,\frac{p^2}{2m}+\Vtrap + U_\eq\Big\}\\
  + \beta \frac{\pv}{m} \cdot \nablav_r (\delta V + \delta U) \Big)
   = -I[\Phi]\,,
\label{eq:BoltzLin}
\end{multline}
where $\{F,G\} = \nablav_r F\cdot \nablav_p G-\nablav_p F \cdot
\nablav_r G$ denotes the Poisson bracket, $\delta V$ is the
perturbation of the trap potential that is used to excite the
collective mode, $\delta U$ is the variation of the mean field due to
the variation of the density, and $I[\Phi]$ is the linearized
collision term. Since we want to calculate the so-called response
function, we take the perturbation to be a pulse,
\begin{equation}
\delta V(\rv,t) = \Vhat(\rv) \delta(t)\,.
\end{equation}
As in \Ref{Chiacchiera2009}, we approximate the variation of the mean field by
\begin{equation}\label{deltaU}
\delta U(\rv,t)= \left.\frac{\partial
  U_\eq}{\partial\rho_\eq}\right|_{\rho_\eq(\rv),T}\, \delta \rho(\rv,t)\, .
\end{equation}
The linearized collision integral reads
\begin{multline}
  I[\Phi] = \int \frac{d^3p_1}{(2\pi)^3} \int d \Omega\,
   \frac{d\sigma}{d\Omega} \frac{|\pv-\pv_1|}{m} \feq f_{\eq 1} 
  \fbareq^\prime \bar{f}_{\eq 1}^\prime\\
    \times (\Phi+\Phi_1-\Phi^\prime-\Phi_1^\prime)\,,
\end{multline}
where an obvious notation for the different $f_\eq$ and $\Phi$ at the
momenta before ($\pv$, $\pv_1$) and after the collision ($\pv^\prime$,
$\pv_1^\prime$) has been used. Note that, especially near the
critical temperature, the in-medium cross-section $d\sigma/d\Omega$
can differ strongly from the free one
\cite{BruunSmith2007,Riedl2008,Chiacchiera2009}.
\subsection{Moments method}
\label{sec:moments}
As in \Ref{Chiacchiera2011}, we are looking for a semi-analytical
solution of the Boltzmann equation (\ref{eq:BoltzLin}) by using the
method of phase-space moments. In this section, we will generalize the
formalism of that paper to the case of an arbitrary trap potential
$\Vtrap$ and with mean field $U$.

The basic idea is to approximate the function $\Phi$ by a polynomial
in the components of $\rv$ and $\pv$ with time-dependent coefficients
$c_j$,
\begin{equation}\label{generalansatz}
  \Phi(\rv,\pv,t)=\sum_{j=1}^n c_j(t) \phi_j(\rv,\pv)\,,
\end{equation}
where the $\phi_j$ are suitable basis functions, e.g., monomials in
the components of $\rv$ and $\pv$.

Multiplying the linearized Boltzmann equation (\ref{eq:BoltzLin}) by
$\phi_i$ and integrating over phase space, one obtains, after a
Fourier transform with respect to $t$, a set of $n$ coupled linear
algebraic equations for the $n$ coefficients $c_j(\omega)$. In
matrix form, they read
\begin{equation}\label{eq:Ac=a}
  \sum_{j=1}^n A_{ij}(\omega) c_j(\omega)= a_i\,,
\end{equation}
where
\begin{gather}
  A_{ij}(\omega) = -i\omega M_{ij} + A^{\trans}_{ij} + A^{\deltaU}_{ij} +
  A^{\coll}_{ij} \,, \label{eq:Aij}
\\
M_{ij} = \int \dgamma \feq\fbareq \phi_i \phi_j\,,
  \label{eq:Mij}
\\
A^\trans_{ij}=-\frac{N}{2\beta}
\ave{\{\phi_i,\phi_j\}}_\eq\,,
\label{eq:Atransij}
\\
A^{\deltaU}_{ij}= \frac{N}{2}\ave{\nablav_p\phi_i\cdot
  \nablav_r\delta U_j}_\eq\,,
\label{eq:AdeltaUij}
\\
A^{\coll}_{ij} = \int \dgamma \phi_i
  I[\phi_j]\,, \label{eq:Acollij}
\end{gather}
and
\begin{equation}
  a_i = - \frac{N}{2}\ave{\nablav_p \phi_i\cdot\nablav_r \Vhat(\rv)}_\eq\,.
\label{eq:ai}
\end{equation}
The variation of the mean field in \Eq{eq:AdeltaUij} is defined as
\begin{equation}
\delta U_j = \frac{\partial U_\eq}{\partial \rho_\eq}
  \int \frac{d^3p}{(2\pi)^3} \feq\fbareq \phi_j\,,
\end{equation}
Equations (\ref{eq:Atransij}) and (\ref{eq:ai}) correspond to Eqs.\@
(14) and (16) in \Ref{Chiacchiera2011} which have been simplified by
integration by parts. Comparing \Eqs{eq:Ac=a}-(\ref{eq:ai}) with the
analogous ones of \Ref{Chiacchiera2011}, Eqs.\@ (11)-(16), one sees
that the mean field gives rise to a new term, $A^\deltaU$. However,
one should keep in mind that implicitly all terms depend on the mean
field since it modifies the equilibrium distribution $\feq$. One can
see that the matrices $M$ and $A^\coll$ are symmetric, $A^\trans$ is
antisymmetric, whereas $A^\deltaU$ has no defined symmetry. In
practice, the calculation of the matrices is straight-forward, but
tedious, and we made use of the Mathematica software to express the
numerous matrix elements in terms of a smaller number of integrals
over equilibrium quantities.

\subsection{Eigenmodes and response function}
\label{sec:response}
Without an external perturbation $\Vhat$, i.e., for $a_i = 0$,
\Eq{eq:Ac=a} has a solution with non-vanishing coefficients $c_j$ only
if $\det A(\omega) = 0$. The frequencies $\omega$ for which this
happens are obviously given by the eigenvalues of the matrix
$-iM^{-1}(A^\trans+A^\deltaU+A^\coll)$. If they are well
separated, it is possible to interpret them as the frequencies of the
eigenmodes of the system. In general, they are complex, and their
imaginary part describes the damping rate of the corresponding mode
\cite{Riedl2008,Chiacchiera2009}.

However, as discussed in \Ref{Chiacchiera2011}, when the moments
method is extended to higher order, there can be many eigenvalues
belonging to a single collective mode. In this case, the scattering of
the eigenvalues, which goes over into a continuous spectrum in the
limit of an infinite number of moments \cite{DaProvidencia1988},
corresponds to a new contribution to the damping (Landau damping) in
addition to the imaginary parts coming from the collision term. In
order to obtain the mode frequency and damping rate in this case, it
is useful to look at the response function which contains the
contributions of all eigenvalues.

We denote by $\ave{q}(\omega)$ the Fourier transform of the
expectation value $\ave{q}(t)$ of some operator $q$ after the
perturbation. The so-called response function is equal to
$\delta\ave{q}(\omega)/\alpha \equiv
(\ave{q}(\omega)-\ave{q}_{eq})/\alpha$ in the special case that the
excitation operator is $\Vhat = \alpha q$. In all excitations
considered in this paper, $\ave{q}_{eq}$ will be
zero. The strength function is proportional to the imaginary part of
the response function.

In order to calculate the response function, we need the coefficients
$c_j(\omega)$ of the ansatz (\ref{generalansatz}). By diagonalizing
the matrix
\begin{equation}
M^{-1}(A^\trans+A^\deltaU+A^\coll)=PDP^{-1}\,,
\end{equation}
with $D=\textrm{diag}(\Gamma_1+i\omega_1,\dots,\Gamma_n+i\omega_n)$,
we can write them as
\begin{equation}
c_j(\omega) = i\sum_{k=1}^{n}
  \frac{P_{jk}(P^{-1}M^{-1}a)_k}{\omega-\omega_k+i\Gamma_k}\,.
\end{equation}
Then, the response function can be easily obtained as $\ave{q}(\omega) =
b^T c(\omega)$, where
\begin{equation}
b_i = \frac{2}{N}\int \dgamma \feq\fbareq q\phi_i\,.
\end{equation}

The frequency and damping of the mode can be extracted from the
response $\ave{q}(\omega)$, e.g., by fitting the peak in the
strength function $-\Im \ave{q}(\omega)$ corresponding to the mode
under consideration with a Lorentzian.

Using the solution for the coefficients $c_j(\omega)$ at the peak, we
can also obtain the velocity field of the corresponding collective
mode:
\begin{equation}
\vv(\rv,\omega) = \sum_{j=1}^{n} c_j(\omega) \frac{1}{\beta\rho_{\eq}}
\int \frac{d^3p}{(2\pi)^3}\feq \nablav_p\phi_j\,.
\end{equation}
\section{Realistic trap potential}\label{sec:potential}
In this section we recall the shape of a typical optical dipole
trap. We concentrate on the configuration used by the Innsbruck group,
consisting of a focused-beam trap \cite{GrimmWeidemueller00} with
additional magnetic confinement in the axial direction.  Following
\Ref{Altmeyer_Thesis}, we parametrize the experimental trap potential
as
\begin{multline}
  \Vexp(\rv) = V_0 \left[1 - \frac{1}{1 + \frac{z^2}{z_0^2}} \exp 
  \left(-\frac{2}{1 + \frac{z^2}{z_0^2} } \frac{\rrad^2}{w_0^2}\right)
  \right]\\
  + \frac{1}{2} m \wzmag^2 z^2 \,, \label{eq:Vtrap}
\end{multline}
where $\rrad = \sqrt{x^2+y^2}$, $V_0$ is the trap depth, $w_0$ is the
minimal waist of the laser beam, $z_0=\pi w_0^2/\lambda$ is the
Rayleigh length, $\lambda$ the laser wavelength and $\wzmag$ defines
the magnetic trapping in the $z$ direction.

\begin{figure}
  \includegraphics[scale=1.2]{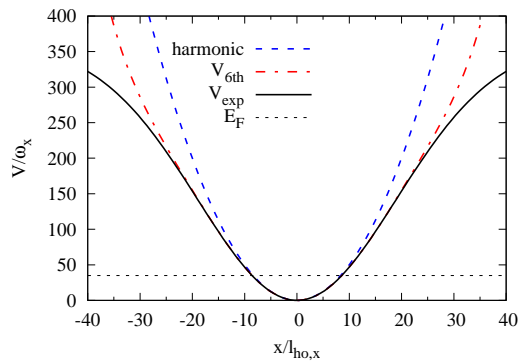}
  \caption{Experimental trap potential (solid line), and its harmonic
    (dashes) and sixth-order approximation (dash-dotted line), as
    functions of $x$ for $y = z = 0$. The Fermi energy for the
    parameters of \Ref{Riedl2008} is indicated by the thin dashes. The
    potentials are in units of $\omega_x$ and $x$ is in units
    $l_{ho,x}=1/\sqrt{m \omega_x}$.}
  \label{fig:vtrap}
\end{figure}

In order to compare our results with the data from
\Refs{Riedl2008,Sidorenkov2012}, we focus on the setup of that
experiment. The trap was of the type (\ref{eq:Vtrap}), with strong
anharmonicity in the $x$-$y$ plane due to the Gaussian shape of the
laser beams. The confinement along the $z$ direction was practically
harmonic. In our calculations, we will expand \Eq{eq:Vtrap} up to
second order in $z$ and up to sixth order in $\rrad$. The result can
be written as
\begin{multline}
\Vsix(\rv) = \frac{m\wrad^2 \rrad^2}{2}
  \Big( 1-\frac{m \wrad^2 \rrad^2}{4 V_0}
    +\frac{m^2 \wrad^4 \rrad^4 }{24 V_0^2}\Big)\\
  + \frac{m \wz^2 z^2}{2}\,,
\label{eq:vsix}
\end{multline}
where $\wrad^2 = 4V_0/(m w_0^2)$ and $\wz^2 = \wzmag^2+\wzopt^2 =
\wzmag^2+2V_0 /(m z_0^2)$ are the radial and axial trap frequencies,
respectively. We use \Eq{eq:vsix} instead of \Eq{eq:Vtrap} because it
simplifies the calculations in the sense that it leads to a density
that tends to zero for $\rrad\to\infty$. The necessity to go beyond
fourth order is also clear from \Eq{eq:vsix}: otherwise the potential
would be unbound from below. For illustration, \Fig{fig:vtrap}
shows the real trap potential (\ref{eq:Vtrap}), its harmonic
approximation, and the one we will use, \Eq{eq:vsix}.

In the present paper, we concentrate on axially symmetric
traps. Nevertheless, we write $\wx$ and $\wy$ instead of $\wrad$ for
the trap frequencies in $x$ and $y$ direction if the formulas can be
generalized to the triaxial case.

In our numerical calculations, we use as an example the parameters of
the Innsbruck experiment \cite{Riedl2008}. As mentioned in
\Ref{Riedl2008}, the sloshing mode was studied with the same
parameters as the compression mode, i.e., $\wrad/(2\pi)=1100$ Hz,
$\wz/(2\pi)=26$ Hz, and $V_0 = 19$ $\mu$K. The trap was loaded with $N
= 6\times 10^5$ atoms of $^6$Li in the unitary limit, $1/(k_Fa)=0$,
and the temperature was varied between $\sim 0$ and $1.2 T_F$. Since
we cannot describe the superfluid phase, we limit ourselves to
temperatures above $0.3 T_F$ \cite{Chiacchiera2009}. Moreover, we
approximate the unitary limit numerically by setting $1/(k_Fa) =
-0.01$.

As pointed out in \Ref{Riedl2008}, the anharmonicity effects depend
mainly upon the ratio $E_F/V_0$. Defining the Fermi energy as usual by
$E_F = (3N\wrad^2\wz)^{1/3}$, one obtains $E_F/V_0 \approx 0.1$, i.e.,
at low temperature the anharmonicity effects are relatively weak. For
illustration, $E_F$ is indicated in \Fig{fig:vtrap} by the thin dashed
line. We see that, for the present choice of parameters, the atoms
start to feel the anharmonicity of the potential when their energy
exceeds $E_F$, while the sixth-order approximation to the potential
stays very precise up to about five times the Fermi energy.
\section{Sloshing mode at first order}\label{sec:firstorder}
In order to describe the sloshing motion, say, along the $x$
direction, the ansatz (\ref{generalansatz}) has to contain at least
two basis functions:
\begin{equation}\label{ansatzfirst}
  \Phi_{\fst}(\rv,\pv,t) = c_1(t) x + c_2(t) p_x\, .
\end{equation} 
The second term describes the collective velocity of the cloud in $x$
direction, while the first one corresponds approximately to a
displacement of the center of mass. The latter statement becomes exact
in the special case of a purely harmonic trap without mean field
($U=0$). In this case, the ansatz (\ref{ansatzfirst}) is closed with
respect to the operator
$\pv/m\cdot\nablav_r-\nablav_r\Vtrap\cdot\nablav_p$ on the left-hand
side of the Boltzmann equation (\ref{eq:BoltzLin}), and since the
collision term does not contribute ($I[x]=I[p_x]=0$ because of
momentum conservation in a collision), the ansatz (\ref{ansatzfirst})
becomes exact. However, if the trap is not purely harmonic, this is no
longer true, since the gradient of the trap potential generates new
terms. And even if the trap is harmonic, it is not true for
interacting atoms, which feel a mean field $U \neq 0$ in addition to
the trap potential $\Vtrap$. In the next section, we will therefore
extend the ansatz (\ref{ansatzfirst}), but let us first look what
happens at first order.

We now apply the formalism described in \Sec{sec:formalism} to
$\Phi_\fst$. The calculation of the matrix $A$ is very easy in this
case. After some algebra, we obtain a quadratic equation for the
frequency of the sloshing mode by imposing $\det A=0$. Its solution
reads
\begin{equation}
  \wslo^2 = \frac{1 - \cpr}{1 - \cpr - \cse} \wx^2\,.
\label{eq:wk1st}
\end{equation}
The parameters
\begin{equation}
  \cpr = \frac{2\beta}{N} \int \dgamma\, \feq \fbareq
  x \frac{\partial U_\eq}{\partial x}
\end{equation}
and
\begin{equation}\label{eq:cse}
  \cse = -\frac{2\beta}{N} \int \dgamma \feq \fbareq
    x \frac{\partial}{\partial x} 
    \Big(\frac{m \wx^2 x^2}{2} - \Vtrap(\rv) \Big)
\end{equation}
characterize, respectively, the strength of the mean field and of the
anharmonicity effects. Equation (\ref{eq:wk1st}) has two important
features: (a) If $\Vtrap$ is harmonic ($\cse = 0$), then $\wslo = \wx$,
independently of the interaction, in accordance with the Kohn theorem
\cite{Kohn}. This point was already discussed in \Ref{Chiacchiera2009}
and shows the consistency of our approach, in particular of
\Eq{deltaU}. (b) If $\Vtrap$ is not purely harmonic ($\cse \neq 0$),
then $\wslo$ depends on both the anharmonicity of the trap and on the
interaction.

Another property of \Eq{eq:wk1st} is that the sloshing mode is
undamped (i.e., $\wslo$ is real) even in the case of an anharmonic
potential. This is, however, only a consequence of the first-order
approximation.

A similar calculation was performed in the Appendix B of
\Ref{Riedl2008}. In that reference, only the leading anharmonicity
correction was kept. If we combine \Eqs{eq:vsix}, (\ref{eq:wk1st}),
and (\ref{eq:cse}) and keep only the leading order in $1/V_0$, we find
\begin{equation}\label{perturbative}
\wslo^2 = \wx^2 \Big(1
  -\frac{m\wrad^2(3\ave{x^2}_\eq+\ave{y^2}_\eq)}{2V_0(1-\cpr)}\Big)\,.
\end{equation}
This result differs from Eq.\@ (B2) of \Ref{Riedl2008} in two
respects. First, there is an additional factor of three in front of
the $\ave{x^2}$ term (misprint in \Ref{Riedl2008}). Second, the
anharmonicity correction is enhanced by a factor $1/(1-\cpr)$. This
factor is missing in \Ref{Riedl2008} because there the mean field was
not considered, although the expectation values $\ave{x^2}$ and
$\ave{y^2}$ were calculated with the density profiles of an
interacting gas.
\section{Sloshing mode at third order}\label{sec:thirdorder}
\subsection{Extended ansatz}\label{sec:extendedansatz}
As in \Ref{Chiacchiera2011}, we will extend the ansatz
(\ref{ansatzfirst}) by including higher-order terms. At the next
higher order, the ansatz contains 18 terms
\begin{equation}\label{ansatzthird}
 \Phi_{\trd}(\rv,\pv,t)=\sum_{i=1}^{18}c_i(t)\phi_i(\rv,\pv)\,,
\end{equation}
where:
\begin{gather}\label{eq:gh}
\phi_{1}=x\,,\, \phi_2=p_x\,,\nonumber\\
\phi_3= x^3 \,,\, \phi_4=x^2 p_x\,, \, \phi_5=x p_x^2 \,,\, \phi_6=p_x^3 \,,\nonumber\\
\phi_7=x y^2\,,\phi_8=y^2 p_x\,,\, \phi_9=x y p_y\,,\, \phi_{10}=y p_x p_y\,\nonumber\\
\phi_{11}= x p_y^2\,,\, \phi_{12}=p_x p_y^2\,,\, \phi_{13}=x z^2\,,\, \phi_{14}=z^2 p_x\,,\nonumber\\
\phi_{15}= x z p_z\,,\, \phi_{16}=z p_x p_z\,,\, \phi_{17}=x p_z^2\,,\, \phi_{18}=p_x p_z^2.
\end{gather}
The first two terms of $\Phi_\trd$ coincide with $\Phi_\fst$, and the
subsequent ones are all possible terms of third order in the
components of $\rv$ and $\pv$ satisfying the symmetries of the mode:
odd under $(x,p_x)\to (-x,-p_x)$ and even under $(y,p_y)\to(-y,-p_y)$
and $(z,p_z)\to(-z,-p_z)$.

Note that some of the third-order terms, e.g., $\phi_4 = x^2 p_x$,
generate a velocity field that is quadratic in the coordinates,
whereas the velocity field corresponding to the first-order ansatz is
spatially constant. Other terms, e.g., $\phi_5 = xp_x^2$ or $\phi_6 =
p_x^3$, describe momentum-sphere distortions. They give rise to a
non-vanishing collision integral so that they generate a damping. The
momentum-independent third-order terms such as $\phi_3 = x^3$ describe
deformations of the cloud in coordinate space that will be discussed
in more detail in the next subsection.

As explained in \Sec{sec:response}, the frequency and damping rate should
now be extracted from the response function. The suitable excitation
operator is
\begin{equation}
\Vhatslo(\rv)=\alpha x\,,
\end{equation}
because it gives all atoms a constant kick in $x$ direction, $p_x\to
p_x-\alpha$, at the moment of the excitation $t=0$. The response
function is defined as the expectation value $\ave{x}(\omega)/\alpha$.

We checked numerically that, in a purely harmonic potential
$(V_0\to\infty)$, the strength function has only a single sharp peak
(no damping) at $\omega=\wx$ within the precision of our
calculation (better than $10^{-4}$), independently of the temperature and
of the interaction strength.

\begin{figure}
\includegraphics[scale=1.2]{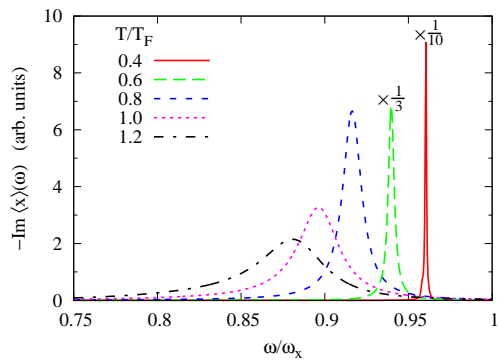}
\caption{Strength function $-\Im\ave{x}(\omega)$ for the
  sloshing mode (excitation $\Vhat = \alpha x$) in the anharmonic trap
  for various temperatures. The system parameters are those of \Ref{Riedl2008}.
  \label{fig:resp_sloshing}}
\end{figure}

In \Fig{fig:resp_sloshing}
we show the strength functions for the sloshing mode in the anharmonic
trap for various temperatures. In all cases the strength is
concentrated in a single peak and the width of the peak corresponds to
the damping rate. The position of the peak is always below
$\omega=\wx$, and with increasing temperature, this shift gets
stronger. This is easily understood: With increasing temperature, more
and more atoms reach the peripheral region where the anharmonic
potential is flatter than the harmonic one (cf. \Fig{fig:vtrap}).

\begin{figure}
\includegraphics[scale=1.2]{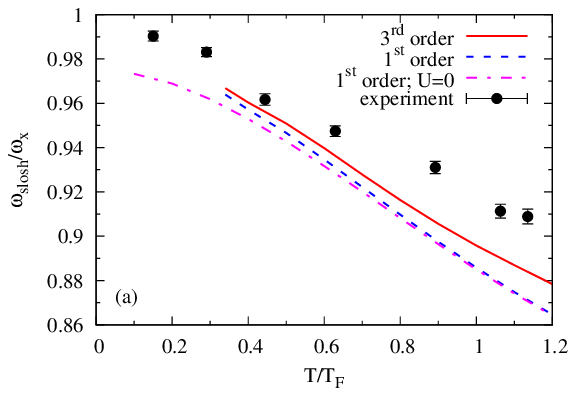}
\includegraphics[scale=1.2]{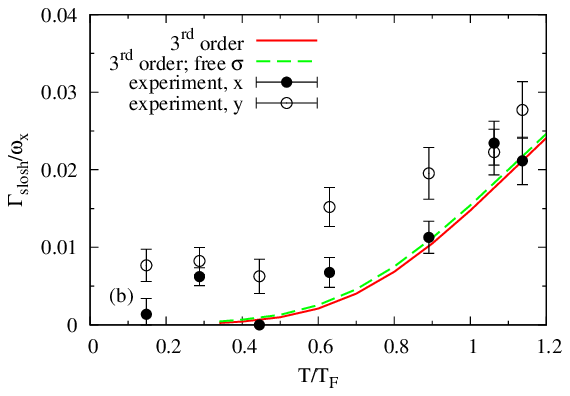}
\caption{(a) Temperature dependence of the frequency of the transverse
  sloshing mode in units of $\wx$. Solid line: third-order result;
  dashed line: first-order result; dash-dotted line: first-order
  result obtained without mean field. The experimental data points are
  taken from \Ref{Riedl2008}. (b) Temperature dependence of the
  damping rate in units of $\wx$. Solid line: full third-order result;
  dashed line: third-order result obtained with the free instead of
  the in-medium cross section. The filled and empty data points are
  measured damping rates of the sloshing modes in $x$ and $y$
  direction, respectively, from \Ref{Sidorenkov2012}.
  \label{fig:sloshing}}
\end{figure}

Since in all the cases shown in \Fig{fig:resp_sloshing} the
strength is concentrated in a single peak, this allows us to extract
the frequency and damping rate by fitting it with a single Lorentzian.
(This is not clear a priori: For instance, if we do the calculation
without mean field, the strength function for low temperature has two
separate peaks, and it is not evident how one has to define the
average frequency and total width.) The results for the frequency and
damping rate as functions of temperature are shown in
\Fig{fig:sloshing}(a)
and (b) as the solid lines. For comparison, the dashed line in
\Fig{fig:sloshing}(a) shows the first-order result for the
frequency. The dash-dotted line corresponds to the first-order result
obtained without mean field. As expected, the mean field reduces the
shift of the frequency, but its effect is moderate and most pronounced
at lower temperature.

The experimental results from Fig.\@ 2 of \Ref{Riedl2008} are also
shown in \Fig{fig:sloshing}(a). In that paper, the measured empirical
temperatures $\tilde{T}$ were already converted into real temperatures
$T/T_F$ as explained in \cite{Kinast2005,Chen2005}, so that the
results can immediately be compared with our calculation. The trend is
correctly reproduced, and the third-order result is clearly in better
agreement with the data than the first-order one, but the theoretical
frequency shift is still stronger than the experimental shifts
reported in \Ref{Riedl2008}. However, one should keep in mind that
while $\omega_\slosh$ was precisely determined, the ratio
$\omega_\slosh/\omega_x$ depends also on $\omega_x$, which was not
very well known in that experiment. Actually, $\omega_x$ was deduced
from $\omega_\slosh$ under the assumption that the anharmonicity
effects become negligible at $T=0$ \cite{Sidorenkov2012}. If we
suppose that the true trap frequency $\omega_x$ was just $\sim
1.5${\%} higher, the data points for $\omega_\slosh/\omega_x$ are
slightly shifted downwards and the agreement between theory and data
becomes much better.

The damping rate, shown in \Fig{fig:sloshing}(b), increases strongly
with the temperature: at $T/T_F = 0.4$ the sloshing mode survives for
several hundreds of oscillation periods, whereas at $T/T_F = 1.2$ its
amplitude decreases by a factor of $1/e$ after $\sim 6$ oscillations.
The in-medium modification of the cross-section has only a small
effect on the damping of the sloshing mode. The experimental data from
\Ref{Sidorenkov2012} are in quite good agreement with our theoretical
result, especially at the high temperature. The strongly scattered
damping rates at low temperature are probably an artefact due to a
beat caused by the small residual ellipticity of the trap potential in
the experiment.

We also compared our results with those shown in Fig.\@ 5(a) of the
recent paper \cite{WuZhang2012_3d} by Wu and Zhang, where they solve
the Boltzmann equation numerically within the relaxation-time
approximation with a local relaxation time $\tau(\rv)$ taken from
\Ref{MassignanBruun2005}. As we discussed in our previous work
\cite{Lepers2010,Chiacchiera2011}, the inclusion of higher-order
moments accounts in an approximate way for the spatial dependence of
the relaxation time. For the sake of comparison, we use the same trap
parameters as Wu and Zhang%
\footnote{The comparison made in Fig.\@ 5(a) of \Ref{WuZhang2012_3d}
  with the data of Fig.\@ 2 of \Ref{Riedl2008} is somewhat misleading
  because the trap parameters are different. In the calculation of
  \Ref{WuZhang2012_3d}, the parameters corresponding to the
  quadrupole-mode measurement of \Ref{Riedl2008} were used, whereas in
  \Ref{Riedl2008} it is said that the sloshing mode was studied with
  the parameters of the compression-mode measurement.}%
, namely, $\wrad/(2\pi)=1800$ Hz, $\wz/(2\pi)=32$ Hz, and $V_0 = 50$
$\mu$K, and neglect, as it is done in \Ref{WuZhang2012_3d}, the mean
field and in-medium effects on the cross section. Because of the
larger trap depth, the frequency shifts are somewhat weaker than those
shown in \Fig{fig:sloshing}(a). Since in the calculation by Wu and
Zhang the frequency depends on the oscillation amplitude, we have to
compare with the results obtained for the smallest amplitude [red
  squares in Fig.\@ 5(a) of \Ref{WuZhang2012_3d}]. If we extrapolate
these results, shown only up to $T/T_F = 0.4$ in \Ref{WuZhang2012_3d},
to higher temperatures, they are in quite good agreement with our
results above $T/T_F \sim 0.5$. At lower temperatures, however, our
frequency shift is slightly stronger than that by Wu and Zhang,
probably because they use Boltzmann instead of Fermi distributions so
that their density is more concentrated in the trap center at low
temperatures.

\subsection{Coupling between sloshing and other modes}
\label{sec:deformation}
\begin{figure}
  \includegraphics[width=4cm]{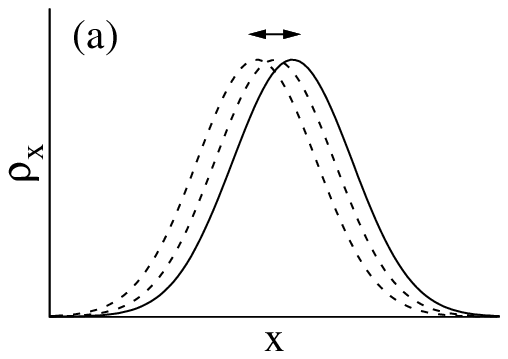}
  \includegraphics[width=4cm]{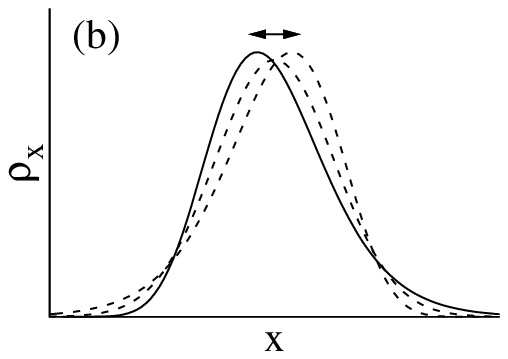}
  \caption{Schematic representation of the oscillation of the density
    profile in $x$ direction ($\rho_x$ being the density integrated
    over $y$ and $z$) for (a) the Kohn mode, in which the cloud
    oscillates as a whole, and (b) the radial compressional dipole
    mode, where the cloud shape oscillates, while the center of mass
    of the cloud stays at rest. All density profiles have the same
    normalization (number of atoms) and the same width.
  \label{fig:sloshskew}}
\end{figure}

\begin{figure}
  \includegraphics[width=4cm]{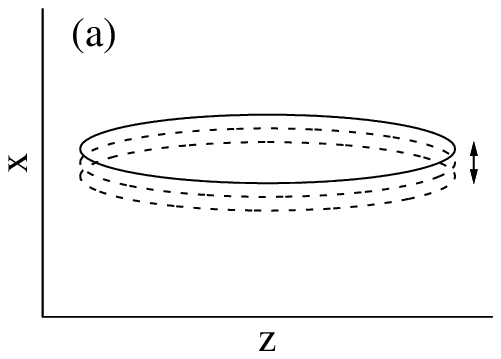}
  \includegraphics[width=4cm]{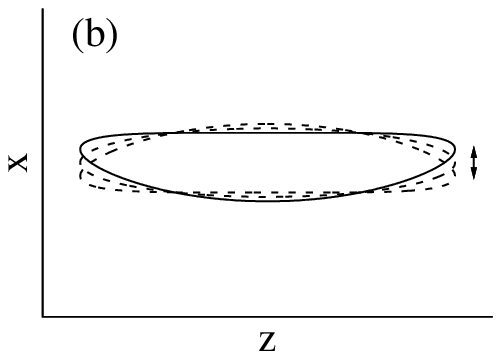}
  \caption{Schematic representation of (a) the Kohn mode, in which the
    cloud oscillates as a whole, and (b) the bending mode, where the
    cloud shape oscillates, while the center of mass of the cloud
    stays at rest. \label{fig:sloshbend}}
\end{figure}

In a harmonic trap, a characteristic feature of the Kohn mode
\cite{Kohn} is that, as long as the interaction is translationally
invariant, the cloud moves as a whole without any change in size or
shape \cite{Dobson}, as illustrated in \Figs{fig:sloshskew}(a)
and \ref{fig:sloshbend}(a).
In an anharmonic trap, this is no longer the case and the cloud will
be slightly distorted during the oscillation. In other words, because
of the anharmonicity, the pure center-of-mass motion gets coupled to
other collective modes. These are damped as usual by collisions, which
results in a damping of the sloshing mode.

What kind of distortions of the cloud shape can one expect? Let us
first note that, to first order in the perturbation, the cloud width
cannot be changed (neither in the transverse nor in the axial
direction), because the excitation operator is odd with respect to
$x\leftrightarrow -x$.

However, the density profile can become skewed in the direction of the
oscillation (i.e., in the $x$ direction in our case). For
illustration, \Fig{fig:sloshskew}(b) shows how the density profile
changes during a radial compressional dipole oscillation. To derive
the corresponding operator, let us start from the usual definition of
the skewness, which is proportional to $\ave{(x-\ave{x})^3}$. Keeping
only terms linear in the variations $\delta\ave{x}$ and
$\delta\ave{x^2}$, and using $\ave{x}_\eq = 0$, we find that for small
oscillations, this is equal to $\ave{x^3-3\ave{x^2}_\eq x}$. If the
trap is axially symmetric, collective modes of different multipolarity
in the radial direction do not mix. It is therefore useful to
decompose the operator into dipolar and octupolar parts:
$x^3-3\ave{x^2}_\eq x = \frac{3}{4} q_\dip + \frac{1}{4} q_\oct$ with
\begin{gather}
q_\dip = (\rrad^2-2\ave{\rrad^2}_{\eq}) \rrad \cos\varphi 
      = (x^2+y^2-2\ave{\rrad^2}_{\eq}) x\,, \label{eq:qdip}\\
q_\oct = \rrad^3 \cos 3\varphi = (x^2-3y^2)x\,. \label{eq:qoct}
\end{gather}
Only $q_\dip$ can get a non-vanishing expectation value during the
sloshing oscillation.

Another possible distortion involves the axial ($z$) direction. Let us
consider again the sloshing mode in $x$ direction. Then one could
imagine that the cloud near $z=0$, owing to its larger radial size, is
more sensitive to the anharmonicity of the trap and therefore
oscillates slightly more slowly than the parts of the cloud at large
$|z|$. Hence, after a few oscillations, the oscillations near $z=0$
and at large $|z|$ become out of phase, which results in a bending of
the cloud in the $x$-$z$ plane, as illustrated in \Fig{fig:sloshbend}(b).
A suitable measure for such a deformation is the expectation value of
the operator
\begin{equation}
q_\bend = (z^2-\ave{z^2}_\eq)x\,. \label{eq:qbend}
\end{equation}
Like $q_\dip$ and $q_\oct$, this operator is defined in such a way
that it is insensitive to a translation of the cloud as a whole in $x$
direction.

\begin{figure}
\includegraphics[scale=1.2]{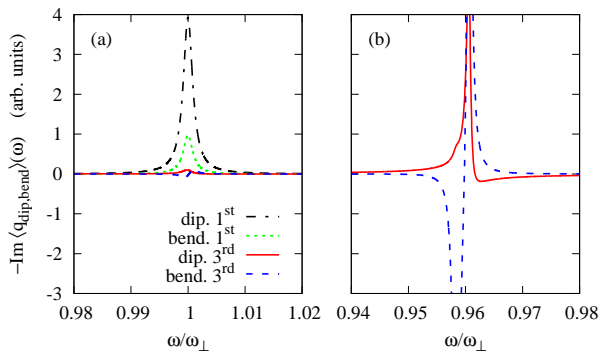}
\caption{Skewness $-\Im\ave{q_\dip}$ and bending $-\Im\ave{q_\bend}$
  of the cloud during the sloshing motion excited by $\Vhat = \alpha
  x$ at $T/T_F = 0.4$. In order to have comparable orders of
  magnitude, we multiplied the bending strengths by a factor
  $(\omega_z/\omega_x)^2$. (a) In a harmonic trap, the unphysical
  skewness and bending of the first-order approximation (dash-dotted
  and dotted lines) are strongly suppressed when the third order
  moments are included (solid and dashed lines). Since in a
  harmonic trap there is no damping, we replaced $\omega$ by $\omega +
  i \, 0.001 \omega_\perp$ to generate this plot. (b) Skewness and
  bending during the sloshing motion in the anharmonic trap within the
  third-order approximation. \label{fig:skewbend}}
\end{figure}

Within the first-order ansatz (\ref{ansatzfirst}), it is easy to see
that in a harmonic trap without mean field and to linear order in the
perturbation, the density oscillates as $\rho(x,y,z,t) =
\rho_\eq(x-\ave{x}(t),y,z)$, with $\ave{x}(t) = c_1(t)T/(m\wx^2)$,
i.e., the cloud shape remains unchanged. However, in the presence of a
mean field $U$, the cloud starts to deform during the oscillation. For
illustration, the measures discussed above for the cloud skewness,
$\ave{q_\dip}$, and for the bending $\ave{q_\bend}$, obtained with the
first-order ansatz in the case of a harmonic trap ($V_0\to\infty$),
are displayed in \Fig{fig:skewbend}(a) by the dash-dotted and dotted lines.
The existence of these distortions is in contradiction to the Kohn
theorem and an unphysical consequence of the crude ansatz. At third
order, owing to the additional degrees of freedom of the ansatz
(\ref{ansatzthird}), the unphysical change of the cloud shape in the
harmonic potential is strongly suppressed, as shown by the solid and
dashed lines in \Fig{fig:skewbend}(a).

While unphysical in a harmonic trap, a change of the cloud shape
during the sloshing motion is expected in an anharmonic trap. From the
preceding discussion it is clear that at least the third-order ansatz
is required to get a meaningful description of this effect. In
\Fig{fig:skewbend}(b), third-order results for the expectation values
of $q_\dip$ and $q_\bend$ during the sloshing motion are displayed.
We note that, compared to the third-order results for a harmonic trap,
both responses are now significantly increased, so that the peaks
correspond to a real physical effect. Moreover, the observed shapes
are typical of what one can obtain in a schematic model by coupling
two damped modes (dipole and bending) to an undamped one
(sloshing). Finally one can also see that the peaks are centered
around the sloshing frequency, which is lower than in the harmonic
case.

\section{Collective modes related to cubic phase-space moments}
\label{sec:cubic}
In the previous section we have discussed coupling of the sloshing
mode to other collective modes due to the trap anharmonicity. In
particular, the ansatz (\ref{ansatzthird}) contains the necessary
terms to describe the transverse compressional dipole and bending
modes corresponding to the operators (\ref{eq:qdip}) and
(\ref{eq:qbend}). In this section, we will study these modes in more
detail. However, one should keep in mind that, since the excitation
operators are already third-order ones, the ansatz (\ref{ansatzthird})
can only give the leading-order result for these modes.

Let us start with the radial dipole oscillation excited by the
operator (\ref{eq:qdip}). In an ideal Fermi gas in an harmonic trap,
this operator would excite two modes with frequencies $\omega/\wrad =
1$ and $3$.

\begin{figure}
\includegraphics[scale=1.2]{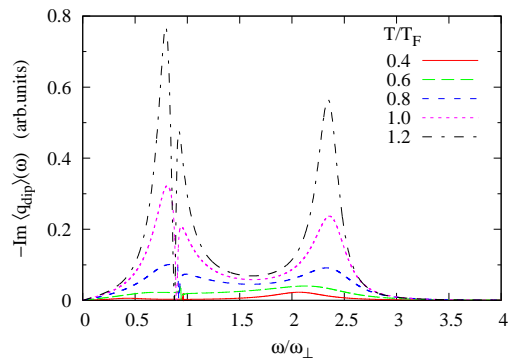}
\caption{Strength function $-\Im\ave{q_\dip}(\omega)$ for the radial
  dipole mode (excitation $\Vhat = \alpha q_\dip$) in the anharmonic
  trap for various temperatures. The parameters are the same as in
  \Fig{fig:sloshing}.
  \label{fig:resp_dip}}
\end{figure}

In \Fig{fig:resp_dip},
the corresponding response function is shown for various
temperatures. We see that in the collisionless regime, i.e., at high
temperature, both modes exist but their frequencies are significantly
lowered by the anharmonicity of the trap. Even at $T/T_F = 1.2$ both
modes are still very strongly damped (damping rate $\sim 0.3
\wrad$). At lower temperature, the damping gets so strong that both
modes disappear completely: the hydrodynamic regime is not reached at
any temperature. By artificially increasing the collision cross
section to reach the hydrodynamic regime at $T/T_F=0.4$, we found that
there would be a mode at $\sim 2 \wrad$ ($\sim 2.4\wrad$ in the case
of a harmonic trap), but in order to reach this regime in an
experiment one would have to use a trap with a much lower radial
frequency (i.e., a more spherical or even pancake-shaped trap). At the
frequency of the sloshing mode, the response exhibits a characteristic
dip due to the coupling between the dipole mode and the almost
undamped sloshing mode (the sloshing is damped mainly because of this
coupling).

\begin{figure}
\includegraphics[scale=1.2]{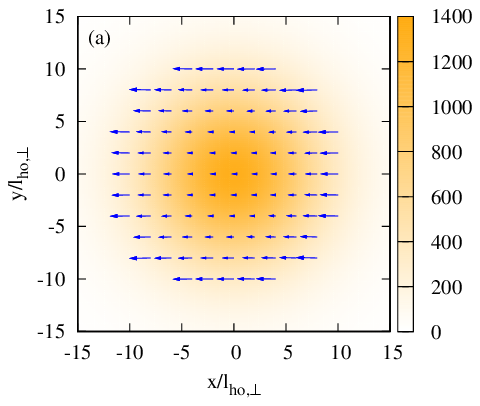}
\includegraphics[scale=1.2]{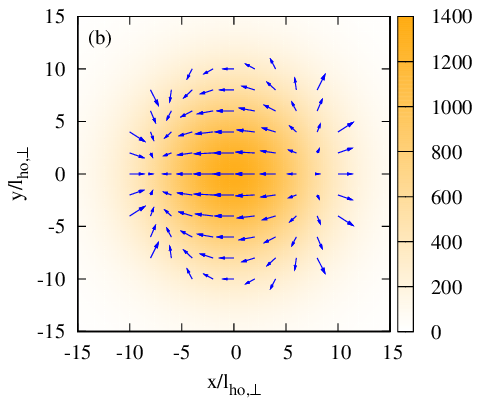}
\caption{Velocity fields in the $x$-$y$ plane, averaged over $z$,
  corresponding to the dipole modes at $\omega \approx 0.8
  \wrad$ (a) and $\omega\approx 2.3\wrad$ (b) at $T=T_F$
  (collisionless regime). The background color indicates the density
  per spin state integrated over $z$ in units of $l_{\ho,\rad}^{-2}$,
  with $l_{\ho,\rad} = 1/\sqrt{m\wrad}$.
  \label{fig:v_dip}}
\end{figure}

In order to get a better understanding of the character of the two
modes in the collisionless regime, we display their velocity fields in
\Fig{fig:v_dip}.
Although both modes are excited by the same operator, their velocity
fields are quite different. The velocity field of the low-lying mode
($\omega\approx 0.8\omega_\perp$ at $T=T_F$), cf.\@  \Fig{fig:v_dip}(a),
confirms that this mode comprises both center-of-mass and skewness
oscillations. The velocity field of the high-lying mode ($\omega
\approx 2.3\wrad$ at $T=T_F$), shown in \Fig{fig:v_dip}(b),
resembles that of the dipole compression modes in atomic nuclei, shown
e.g., in Fig.\@ 6 of \Ref{Serr}.

\begin{figure}
\includegraphics[scale=1.2]{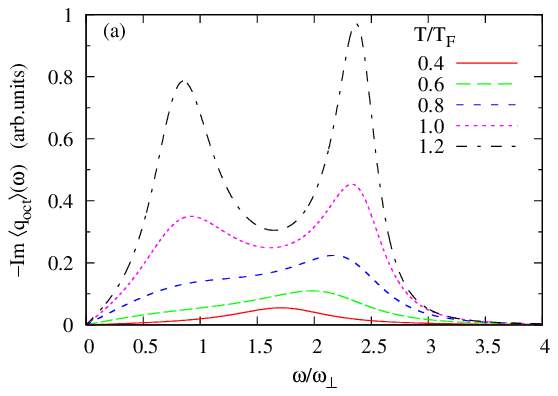}
\includegraphics[scale=1.2]{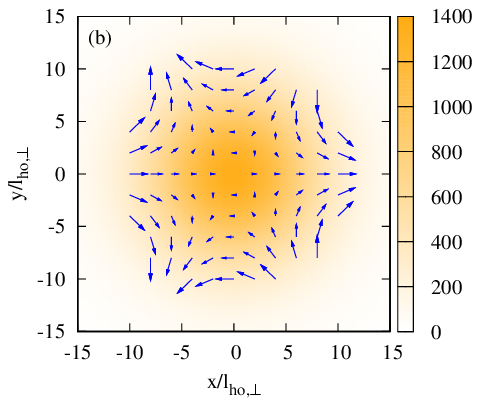}
\caption{(a) Strength function $-\Im\ave{q_\oct}(\omega)$ for the
  radial octupole mode (excitation $\Vhat = \alpha q_\oct$) in the
  anharmonic trap for various temperatures. The parameters are the
  same as in \Fig{fig:sloshing}. (b) Velocity field corresponding to
  the high-lying mode at $T = T_F$ (see caption of \Fig{fig:v_dip} for
  details).
  \label{fig:octupole}}
\end{figure}

Although in an axially symmetric trap the octupole mode does not
couple to the sloshing mode, let us for the sake of completeness
briefly discuss this mode. Like the dipole operator (\ref{eq:qdip}),
the octupole operator (\ref{eq:qoct}) would excite two modes at
$\omega/\wrad = 1$ and $3$ if the gas was collisionless and in a
harmonic trap. The octupole response function for the realistic case
is displayed in \Fig{fig:octupole}(a).
As in the dipole response, both modes are considerably shifted
downwards because of the trap anharmonicity. The damping is even
stronger than in the case of the dipole modes. Both octupole modes
have a similar velocity field, as an example \Fig{fig:octupole}(b)
shows that of the higher-lying mode. In a nuclear physics context it
was found that their difference lies mainly in the non-diagonal
pressure tensor (quadrupole moments in momentum space)
\cite{KohlSchuck}. At $T/T_F=0.4$, there is only a very broad peak
around the hydrodynamic frequency $\omega/\wrad = \sqrt{3}$ predicted
in \Ref{GriffinWu1997} for a harmonic trap. However the strong damping
shows that one is still far from the hydrodynamic regime. By
increasing again artificially the cross-section, the hydrodynamic
frequency is found to be lowered to $\simeq 1.58 \wrad$ due to
anharmonicity.

\begin{figure}
\includegraphics[scale=1.2]{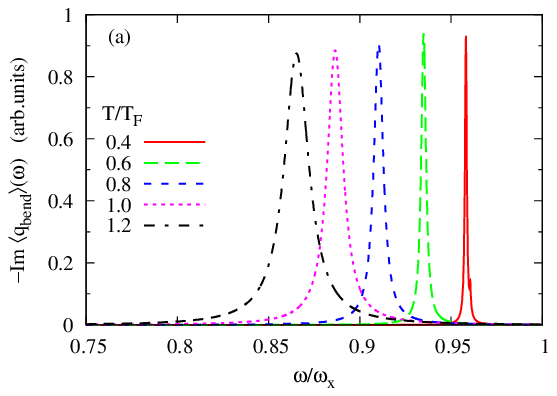}
\includegraphics[scale=1.2]{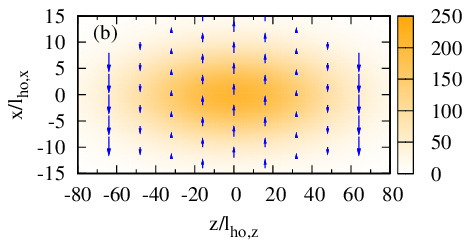}
\caption{(a) Strength function $-\Im\ave{q_\bend}(\omega)$ for the
  bending mode (excitation $\Vhat = \alpha q_\bend$) in the anharmonic
  trap for various temperatures. The parameters are the same as in
  \Fig{fig:sloshing}. (b) Velocity field in the $x$-$z$ plane,
  averaged over $y$, corresponding to the bending mode at $T =
  T_F$. The background color indicates the density integrated over $y$
  in units of $1/(l_{\ho,x}l_{\ho,z})$ with
  $l_{\ho,i}=1/\sqrt{m\omega_i}$.
  \label{fig:bending}}
\end{figure}

Finally, in \Fig{fig:bending}
we display the strength function and the velocity field of the bending
mode. In a non-interacting gas in a harmonic trap, there would be
three modes at $\omega = \omega_x$ and $\omega =
\omega_x\pm2\omega_z$. In the realistic case, these peaks cannot be
resolved and one sees a damped oscillation with a frequency close to
$\omega_x$. In fact, the behavior of its frequency and damping rate is
qualitatively similar to that of the sloshing mode, but the
temperature dependence of the frequency is stronger. Above $0.5 T_F$,
the damping is weaker than that of the sloshing mode, but this is
probably an artefact of our approximation: We expect that the bending
mode will receive an additional damping (analogous to that of the
sloshing mode) from its coupling to a radial dipole mode modulated in
$z$ [operator $(z^2-\ave{z^2}_\eq)q_\dip$], which is not included in the
third-order ansatz. Except at the lowest temperature, the mixing
between sloshing and bending mode is weaker than that between sloshing
and dipole mode, and the bending response does not show the
characteristic dip that was found in the dipole response. The velocity
field, shown in \Fig{fig:bending}(b) confirms that this mode
corresponds to a bending of the elongated cloud in the $x$-$z$ plane.

\section{\label{sec:cl}Conclusions}
In this paper, we studied the frequency shift and damping of the
sloshing mode of a normal-fluid Fermi gas in a realistic anharmonic
trap potential. We used the moments method in order to find
approximate solutions of the Boltzmann equation with mean field and
in-medium cross-section. Already at first order, the moments method
predicts a downwards shift of the frequency of the sloshing
mode. However, the first order ansatz is insufficient to describe the
coupling between the center-of-mass and the internal degrees of
freedom of the gas, which is responsible for the damping of the
sloshing mode in an anharmonic trap. We therefore extended the ansatz
to include also third-order phase-space moments. The third-order
ansatz contains not only the sloshing mode, but also more complicated
radial dipole and bending modes. The important mechanism for the
damping of the sloshing mode is its coupling to these modes
(especially to the dipole mode) due to the trap anharmonicity. In
addition, the third-order ansatz allowed us to discuss the radial
octupole mode.

The comparison with the available experimental data for the sloshing
mode shows that the essential features of the frequency shift can be
reproduced. Quantitatively, the calculated ratios
$\omega_\slosh/\omega_x$ are somewhat lower than those given in
\Ref{Riedl2008}. Since both the mean field and the third-order moments
give only small corrections to the frequency shift predicted by the
first-order moments method, further improvements of the theory will
probably give even smaller corrections. An effect which has not been
considered in our linear-response study is the dependence of the
sloshing frequency on the amplitude of the mode. However, this effect
goes into the wrong direction, since the restoring force gets weaker
with increasing amplitude (see \Fig{fig:vtrap}) and therefore the
frequency of the sloshing mode will be further reduced (cf. also
\cite{WuZhang2012_3d}). As discussed in \Sec{sec:extendedansatz}, a
possible explanation of the discrepancy is that the experimental trap
frequency $\omega_x$ may have been slightly underestimated in the
analysis of \Ref{Riedl2008}. Actually, our results could be used to
determine the true value of $\omega_x$.

The agreement between the calculated and measured damping rates of the
sloshing mode is satisfactory and makes us confident that our approach
contains the essential physics to describe the coupling between the
sloshing and internal degrees of freedom. This may also help to better
understand other effects where the dissipation of center-of-mass
kinetic energy plays a role, such as the heating of the cloud due to
laser-beam-pointing noise discussed in \Ref{SavardOHara97}.

In addition to the sloshing mode, we studied the radial dipole and
octupole modes and the bending mode in the $x$-$z$ plane.  However,
one should keep in mind that for these modes the third-order ansatz
contains only the leading order, and higher-order corrections can be
important. We found that, with the present trap parameters, the dipole
and octupole modes do not behave hydrodynamically at any
temperature. The frequencies of the dipole, octupole, and bending
modes are more strongly affected by the anharmonicity than the
frequency of the sloshing mode. For the dipole and octupole modes,
this can be intuitively understood, since these higher-order modes are
more sensitive to what happens at larger $r_\perp$, where the atoms
feel the anharmonicity of the trap. The same argument should also
apply to the radial quadrupole and scissors modes not studied in the
present paper. One should therefore be careful when comparing
theoretical results obtained for a harmonic trap with experimental
ones (as done in \cite{Riedl2008,Chiacchiera2009,Chiacchiera2011}),
since the anharmonicity effects cannot be completely absorbed in a
renormalized radial trap frequency.

As a final remark we note that a similar third-order ansatz as the one
used here [\Eq{ansatzthird} with $(x,p_x)\leftrightarrow(z,p_z)$] can
be used to describe one of the higher-order axial modes studied
recently by the Innsbruck group \cite{Sanchez_Thesis} (the $k=2$ mode
in the notation of \cite{Sanchez_Thesis}, corresponding to a motion as
shown in \Fig{fig:sloshskew}(b) but in $z$ instead of $x$ direction).
Having more nodes than the usually considered collective modes, these
kinds of modes are very interesting since they are closer to sound
waves in uniform systems.
\section*{Acknowledgements}
We thank L. Sidorenkov and R. Grimm for useful explanations concerning
the experimental data of \cite{Riedl2008,Sidorenkov2012} and for the
permission to use the unpublished data of
\cite{Sidorenkov2012}. S.C. is supported by the \emph{Funda\c{c}\~ao
  para a Ci\^encia e a Tecnologia} (FCT, Portugal) and the
\emph{European Social Fund} (ESF) via the post-doctoral grant
SFRH/BPD/64405/2009.  S.C. also acknowledges partial support by
QREN/FEDER, the COMPETE Programme, and FCT under the project No.
PTDC/FIS/113292/2009.

\end{document}